\documentclass[
 reprint,
 amsmath,amssymb,
 prx,showpacs,superscriptaddress,aps
]{revtex4-2}

\usepackage{graphicx}
\usepackage{dcolumn}
\usepackage{bm}
\usepackage{hyperref}
\usepackage{xcolor}
\usepackage{amsmath}
\usepackage[onehalfspacing]{setspace}
\begin{document}
\newcommand{\gket}{$|g\rangle$}
\newcommand{\aket}{$|a\rangle$}
\newcommand{\bket}{$|b\rangle$}
\newcommand{\cket}{$|c\rangle$}
\newcommand{\dket}{$|d\rangle$}
\newcommand{\eket}{$|e\rangle$}
\newcommand{\fket}{$|f\rangle$}
\newcommand{\cm}{cm$^{-1}$}

\setstretch{1.1}

\title{Magnetic atoms with a large electric dipole moment}

\author{J. Seifert}
\affiliation{%
 Fritz-Haber-Institut der Max-Planck-Gesellschaft, Faradayweg 4-6, 14195 Berlin, Germany.
}
\author{S. C. Wright}
\affiliation{%
 Fritz-Haber-Institut der Max-Planck-Gesellschaft, Faradayweg 4-6, 14195 Berlin, Germany.
}
\author{B. G. Sartakov}
\affiliation{%
 Fritz-Haber-Institut der Max-Planck-Gesellschaft, Faradayweg 4-6, 14195 Berlin, Germany.
}
\author{G. Valtolina}%
\email{valtolina@fhi-berlin.mpg.de}
\affiliation{%
 Fritz-Haber-Institut der Max-Planck-Gesellschaft, Faradayweg 4-6, 14195 Berlin, Germany.
 }
\author{G. Meijer}%
\email{meijer@fhi-berlin.mpg.de}
\affiliation{%
 Fritz-Haber-Institut der Max-Planck-Gesellschaft, Faradayweg 4-6, 14195 Berlin, Germany.
}%

\date{\today}

\begin{abstract}
We experimentally show that an electric dipole moment of more than 1 Debye can be induced in the dysprosium (Dy) atom, in a long-lived state that is about 17513 cm$^{-1}$ above the ground state. This metastable state is part of a strongly coupled opposite-parity doublet. Using optically detected microwave spectroscopy in an atomic beam, we determine the approximately 1.12 cm$^{-1}$ doublet spacing for the five stable bosonic isotopes of Dy with kHz-level accuracy. From the shift of the microwave transition frequency in low electric fields (below 150 V/cm) and from optical spectra in high electric fields (up to 150 kV/cm), a reduced transition dipole moment of 7.65 $\pm$ 0.05 Debye between the doublet states is extracted. In high electric fields the doublet interacts with a third state at 17727 cm$^{-1}$, that connects to the ground state via an electric-dipole transition. The three-state Stark interaction enables preparation of Dy atoms in the metastable state via single-photon excitation from the ground state. \textcolor{black}{Our work experimentally validates predicted characteristics of the opposite-parity doublet, and provides an effective route to the first doubly polar quantum gas of atoms.}
\end{abstract}

\maketitle


 
The introduction of laboratory-frame dipole moments in atomic and molecular systems is associated with rich and interesting physics. In particular, long-range and anisotropic dipole-dipole interactions enable the realization of new phases of matter \cite{Lahaye2009, Baranov2012} and deterministic entanglement generation for quantum computing \cite{Holland2023,Bao2023, Picard2025, Ruttley2025}. There is a growing interest in systems for which both large electric and magnetic dipole moments are available, since this allows for an interplay of interactions that can be tuned with independent static fields \cite{Micheli2006, Finelli2024}.

Whilst large magnetic dipole moments are inherent to many species due to the intrinsic spin of the electron, electric dipole moments arise through the interaction of opposite parity levels when an external electric field is applied. In heteronuclear, diatomic molecules, rotational motion of the nuclei naturally gives rise to opposite parity levels separated by one unit of angular momentum, with spacings on the order of 1~cm$^{-1}$; with the inclusion of electronic orbital angular momentum, more closely spaced opposite parity levels with equal total angular momentum ($\Omega$-doublets) are generally present. This can enable large laboratory-frame electric dipole moments, with important applications in manipulating and control of neutral molecules in general \cite{Meerakker2012} and in testing fundamental physics \cite{Andreev2018,Roussy2023} in particular.

The more symmetric structure in atoms generally precludes the generation of large electric dipole moments, except in the Rydberg regime, where, however, the atomic size and density of electronic states diverges. The dysprosium (Dy) atom represents the textbook exception. In its ground state, Dy has an even-parity electronic configuration [Xe]$4f^{10}6s^{2}$, with [Xe] representing the electronic configuration of the xenon atom. The $4f$-electrons combine into spin and orbital angular momentum quantum numbers of $S=2$ and $L=6$, respectively, and form a $^5I$ state with five $J$-levels. In the $J$=8 ground state, Dy has one of the largest magnetic dipole moments in the periodic table, which has recently allowed the realization of the long-sought-after dipolar supersolid phase \cite{Chomaz2023, Boettcher2019, Tanzi2019, Chomaz2019}. The presence of different high-$J$ levels in Dy, combined with the large energy spacing from spin-orbit effects, makes its electronic spectrum rather dense. Many excited levels are nearly degenerate, and there are opposite parity doublets whose total angular momentum is the same or differs by one. One such doublet, lying about 19798 cm$^{-1}$ above the ground state, with a splitting in the (sub-) GHz range, has been used to search for time-variation of the fine structure constant \cite{Budker1993,Roberts2015}. However the radiative lifetime of both levels is short (8 $\mu$s and 130 $\mu$s), and the maximum induced electric dipole moment is small (0.006 D), making its application in dipolar physics limited.
\begin{figure*}
\includegraphics[]{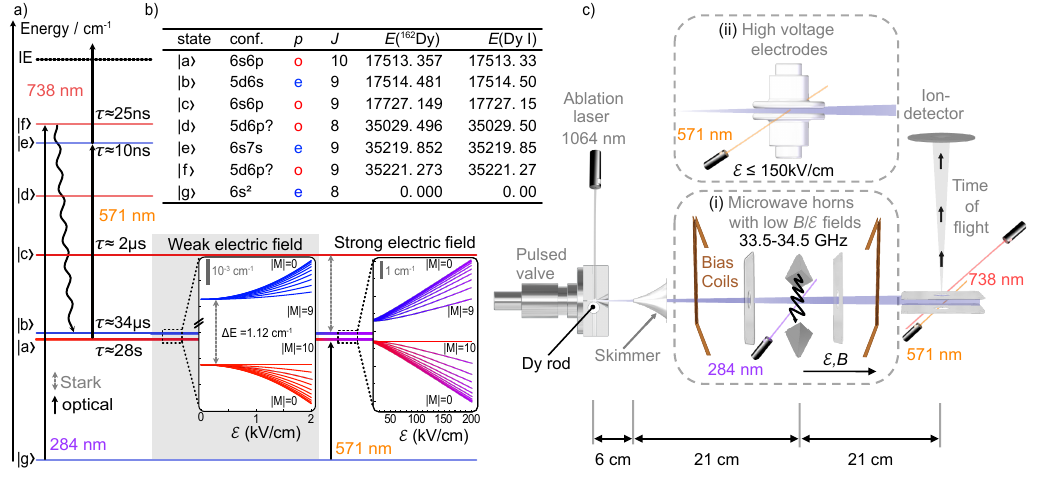}
\caption{(a) Selected energy levels of Dy relevant for this study (not to scale). The quadratic (linear) shift of the $M$-levels of the opposite-parity doublet in weak (strong) electric fields is shown in the left (right) inset. (b) The labeling (state), characteristics (electronic configuration (conf.); parity (p), even (e) or odd(o); total electronic angular momentum ($J$)) and measured energies (E, for $^{162}$Dy, in cm$^{-1}$) of the states are given. In the configuration the common term [Xe]$4f^{10}$ has been omitted. The error bar in the measured energies is $\pm$ 0.002 cm$^{-1}$. The electronic configurations and the energies in the last column are from the NBS Handbook \cite{NBS1978}, i.e. for $^{162}$Dy. (c) Scheme of the atomic beam machine. Dy atoms are produced by ablation of a rotating Dy rod with pulsed 1064 nm radiation and expand, seeded in a pulse of rare gas, into the source chamber. After the skimmer, in a compact interaction chamber, (i) optical and microwave transitions can be induced in the presence of weak magnetic and electric fields, or (ii) the atomic beam can be exposed to electric fields up to 150 kV/cm. In the detection chamber the Dy atoms are ionized via various REMPI schemes and the ions are mass-selectively detected in a time-of-flight mass spectrometer.}
\label{Fig_ExperimentalDetails}
\end{figure*}

In this Letter, we experimentally characterize the opposite-parity doublet whose Stark interaction was noted in passing by Wyart in Dy emission spectra over 50 years ago \cite{Wyart1974}. This doublet is composed of a lower $J$=10 level and an upper $J$=9 level, separated by slightly over 1 cm$^{-1}$, lying around 17513 cm$^{-1}$ above the ground state. Under field-free conditions, the $J$=10 level lives about half a minute and a large transition dipole moment of 8.16 D between the doublet states has been predicted \cite{Lepers2018}. This system is of considerable interest as it offers a platform for quantum simulation of new many-body phases with two independent and competing long-range dipolar interactions \cite{Mishra2020,Ghosh2022, Anich2024, Ghosh2024}, without the challenges posed by ultracold molecules \cite{Langen2024}. 

Following the convention of Ref.~\citenum{Lepers2018}, we label the lowest state of the doublet as \aket\ and the upper one as \bket, see Fig.~\ref{Fig_ExperimentalDetails}. Direct electric dipole transitions to either one of these states from the ground state, labeled as \gket, are forbidden. The \aket\ state has no electric-dipole allowed transition to any lower lying state and its calculated radiative lifetime is $\tau_a$ = 28.1 s. For \bket\ a lifetime of $\tau_b$ = 34 $\mu$s is predicted \cite{Lepers2018}.

We set up an atomic beam experiment to accurately determine the absolute energies of \aket\ and \bket\ for $^{162}$Dy and $^{164}$Dy, the two most abundant bosonic isotopes, to ease the search for the microwave transition between these states. Ablation of a Dy rod (natural isotopic abundance) with a pulsed Nd:YAG laser in a Smalley-type source is used to create a pulsed, supersonic jet (10 Hz repetition rate) of Dy atoms seeded in either helium or argon gas. Using a skimmer with a 4 mm diameter opening, a collimated beam is formed that passes through a short interaction chamber before it enters a differentially pumped detection chamber. In the interaction chamber, laser radiation and microwaves can be made to interact with the Dy atoms in the beam in the presence of weak magnetic and electric fields. Two aluminum plates separated by 28 mm are used to generate electric fields up to 150 V/cm, directed along the molecular beam axis; a 5 mm diameter hole at the center of the plates allows the atoms to pass through. Sets of coils in Helmholtz configuration outside the chamber produce a homogeneous magnetic field up to 10 G parallel to the electric field. Alternatively, the atomic beam can be made to pass in between a pair of polished, high voltage electrodes spaced by 2.020(10) mm. In this setup, described elsewhere \cite{Truppe2019}, we can apply fields  up to 150 kV/cm, see Fig.~\ref{Fig_ExperimentalDetails}. In the detection chamber, various resonance-enhanced multiphoton ionization (REMPI) schemes are used. Ion detection after time-of-flight mass separation then enables fully isotope-resolved recording of the spectra. 

\begin{figure}[t!]
\includegraphics[scale=0.55]{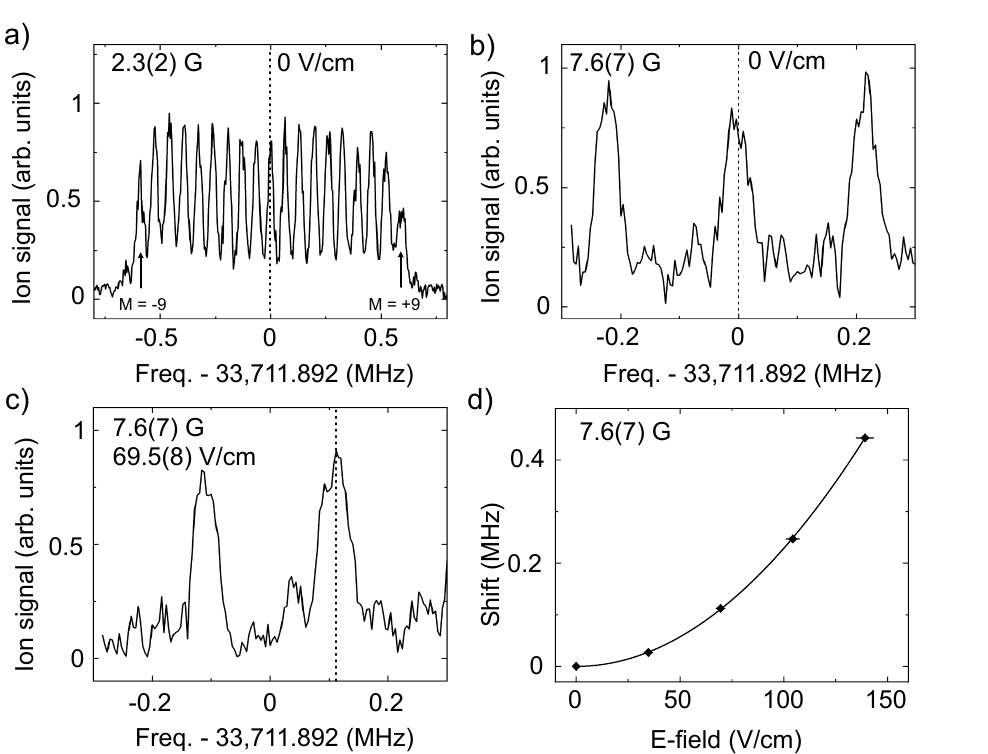}
\caption{Ionization detected microwave spectra of the $|b\rangle \rightarrow |a\rangle$ transition for $^{162}$Dy in different magnetic and electric fields. The vertical dashed line indicates the $M_b=0 \rightarrow M_a=0$ transition. a) The $\Delta M=0$ branch in a magnetic field of 2.3 Gauss. b) At $B$=7.6 Gauss, the magnetic-field insensitive $M_b=0 \rightarrow M_a=0$ transition is fully isolated, allowing for the precise determination of $\Delta E$. c) When a weak electric field is added, the $M_b=0 \rightarrow M_a=0$ transition is up-shifted. d) Measured up-shift of the $M_b=0 \rightarrow M_a=0$ transition as a function of the applied electric field.} 
\label{Fig_MW_peaks}
\end{figure}
The table in Fig.~\ref{Fig_ExperimentalDetails} shows our measured energies of six electronic states of $^{162}$Dy, labelled \aket\ through \fket. To determine these, we scan a narrowband pulsed dye amplified (PDA) laser system, injection seeded with a cw dye laser that is calibrated with a wavemeter, in the 564-572 nm range. The PDA has a spectral bandwidth of about 0.007 cm$^{-1}$ at the fundamental wavelength, and either the fundamental or the frequency-doubled output of this laser is used. The PDA beam crosses the atomic beam perpendicularly in the ionization chamber. There, it is overlapped with the counter-propagating beams of a pulsed dye laser (bandwidth 0.05 cm$^{-1}$) or a second PDA, at a fixed frequency and at the appropriate time, used either for preparation of the Dy atoms in a certain state or for ionization; efficient ionization from \eket\ and \fket\ is performed by exciting to a broad auto-ionizing resonance with 738 nm light from the second PDA. The energies of \cket, \dket, and \fket\ are determined by scanning the PDA over the corresponding transitions from \gket. The \eket\ state energy is determined by scanning the \eket $\leftarrow$\cket\ transition with the PDA after laser preparation of the atoms in \cket. When using helium as a carrier gas, electronic quenching in the expansion is incomplete and a sufficient number of Dy atoms populate \aket\ in the beam, that the \eket $\leftarrow$\aket\ transition can be recorded to determine the energy of \aket. After \fket $\leftarrow$\gket\ excitation with pulsed UV-radiation, \bket\ is populated via spontaneous fluorescence \cite{Wyart1974}. By subsequently scanning the \fket $\leftarrow$\bket\ transition with the PDA, the energy of \bket\ is determined. 
\textcolor{black}{Although we cannot determine the lifetime of \bket\ precisely, our delayed ionization measurements show the lifetime to be in the 25-45 $\mu$s range.} The energies given in the Table in Fig.~\ref{Fig_ExperimentalDetails} are accurate to better than $\pm$ 0.002 cm$^{-1}$. The energies for \aket\ and \bket\ differ from the values in the NBS Handbook, such that $\Delta E$ $\equiv$ $E_b$ - $E_a$ deviates by 0.05 cm$^{-1}$ from the NBS value \cite{NBS1978}.
\begin{figure*}
\includegraphics[scale=0.7]{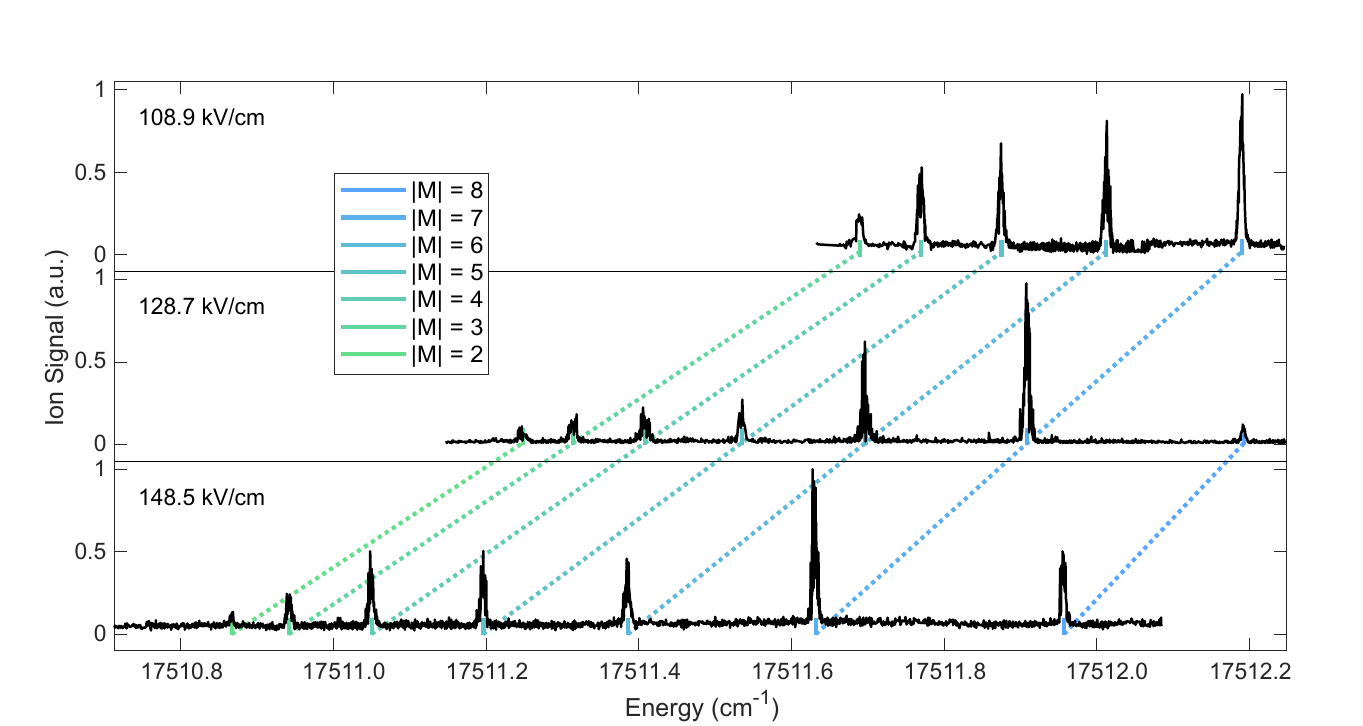}
\caption{\textcolor{black}{Measured excitation spectrum from \gket\ to the M-components of \aket\ in an electric field of 108.9 kV/cm (top), 128.7 kV/cm (center), and 148.5 kV/cm (bottom) for $^{164}$Dy. The spectra are composed of multiple scans over about 0.2 cm$^{-1}$ wide, overlapping intervals. The colored dashed lines connect the centers of the peaks for the same $|M|$-components at different fields. The Stark shift for $|M|\leq$4 is consistent with an EDM larger than 1 D.}}
\label{Fig_StarkJ10}
\end{figure*}

To accurately measure the energy separation between the doublet states, we prepare the Dy atoms in the interaction chamber in \bket\ as described above. We then couple a 20 $\mu$s duration pulse of freely propagating microwave radiation from a horn antenna into the interaction chamber, perpendicular to the molecular beam, to transfer the atoms to \aket. We use argon as a carrier gas, both to reduce the amount of background atoms in \aket\ in the beam and to reduce the speed of the atoms to about 600 m/s, so that they only travel about 12 mm during the application of the microwave pulse. Over this distance, the applied weak magnetic and electric fields are constant to within a few percent. The Dy atoms in \aket\ take about 350 $\mu$s to reach the detection region, where they are ionized via the \eket $\leftarrow$\aket\ transition. 

\begin{table}[b!]
\begin{center}
\begin{tabular}{ c c c c | c } 
\hline
           &$\Delta E$ [MHz] ~&  IS [MHz]   ~&    rIS  ~&   rFS \\
\hline 
$^{156}$Dy & 34,327.948(30)  ~& 776.779(30) ~&  4.8330 ~&  ~4.849 \\
$^{158}$Dy & 34,059.154(30)  ~& 507.985(30) ~&  3.1606 ~&  ~3.167 \\
$^{160}$Dy & 33,887.055(3)   ~& 335.886(4)  ~&  2.0898 ~&  ~2.093 \\
$^{162}$Dy & 33,711.892(3)   ~& 160.723(4)  ~&  1.0000 ~&  ~1.000 \\
$^{164}$Dy & 33,551.169(3)   ~&   0.000     ~&  0.0000 ~&  ~0.000 \\
\hline
 \end{tabular}
 \end{center} 
\caption{Measured values, with error bars, of $\Delta$E $\equiv$ E$_b$ - E$_a$ for the five stable bosonic isotopes of Dy as determined from the microwave spectra. The Isotope Shift (IS) relative to the value for $^{164}$Dy as well as the relative Isotope Shift (rIS; setting IS($^{162}$Dy) $\equiv$ 1.0 ) are given. Interestingly, the latter compare well to the relative Field Shift (rFS) values reported for Dy in Ref.~\citenum{Dekker1968}, given in the last column.}\label{TableMicrowave}
\end{table}

In an external magnetic field $B$, the $M_i$-levels of the opposite-parity doublet experience a Zeeman shift and their energies are given by $E_i + g_i\cdot\mu_B B\cdot M_i$, with $i$ referring to either \aket\ or \bket, $\mu_B$ being the Bohr magneton, and $g_i$ the Land\'e $g$-factors. The $\Delta M$=0 branch of the ionization detected \bket $\rightarrow$\aket\ microwave transition of $^{162}$Dy is shown in Fig. \ref{Fig_MW_peaks}(a). The UV-radiation that is used to drive the \fket $\leftarrow$\gket\ transition is linearly polarized, perpendicular to the applied magnetic field. Assuming an equal population for all 17 $M_g$-levels and in the absence of saturation on the UV transition, this results, after spontaneous fluorescence, in a population distribution among the $M_b$-levels that is proportional to 1.0 + (29/2190)$\cdot$$M_b^2$. The transition intensities of the lines of the $\Delta M$=0 branch of the \bket $\rightarrow$\aket\ transition are proportional to (100 - $M_b^2)$. The product of these two factors yields the rather flat $M$-level intensity distribution for the 19 lines observed in the experiment. The central, magnetic field insensitive $M_b=0 \rightarrow M_a=0$ transition, indicated by a vertical dashed line in Fig. \ref{Fig_MW_peaks}(a-c), yields the precise value for $\Delta E$. The resolved microwave peaks have a width of about 45 kHz, allowing us to determine $\Delta E$ with a precision of few kHz. The thus determined values of $\Delta E$ for $^{160}$Dy, $^{162}$Dy and $^{164}$Dy are shown in Table \ref{TableMicrowave}. For the low abundance $^{156}$Dy and $^{158}$Dy isotopes the individual $M$-components are not resolved, resulting in a larger error bar. In the literature, the Land\'e $g$-factors are given as $g_a$=1.30 and $g_b$=1.316 \cite{NBS1978}; our full microwave spectra indicate that $g_b - g_a$ = 0.025(2).

In an electric field $\mathcal{E}$ the $M$-levels of \aket\ and \bket\ interact and repel each other. When the electric and magnetic fields are parallel, the energies of the $M$-levels are given by \cite{Lepers2018}
\begin{equation}
\frac{E_a+E_b}{2} \pm \sqrt{\left(\frac{\Delta E}{2}\right)^2 + \frac{100 - M^2}{3990}\cdot (\mu_\text{TDM} \mathcal{E})^2}
\end{equation}
where the + (-) is for \bket\ (\aket) and where $\mu_\text{TDM}$ is the reduced transition dipole moment between \aket\ and \bket. In the weak-field limit, the magnetic field insensitive microwave transition will then be up-shifted by $\delta \nu$ = (20/399)$\cdot \left(\mu_\text{TDM} \cdot \mathcal{E}\right)^2/\Delta E$. The experimental data shown Fig. \ref{Fig_MW_peaks}(d) show a purely quadratic dependence on the electric field strength with $\delta \nu$ = 22.8(7)$\cdot (\mathcal{E}\textrm{[V/cm]})^2$ Hz for $^{162}$Dy. If interpreted as solely being due to the Stark interaction (see Supplementary Information) then this implies a value of $\mu_\text{TDM}$ = 7.8 $\pm$ 0.2 D. 

The (main) electronic configuration and parity of \cket\ are the same as those of \aket. It is expected, therefore, that the transition dipole moment $\mu^{\prime}_\text{TDM}$ between \cket\ and \bket\ is also quite large. Even though the spacing $\Delta E^{\prime}$ between these states is almost a factor 200 larger than $\Delta E$, a three-level Stark interaction is expected to become noticeable in high electric fields. This is particularly interesting, as the \cket $\leftarrow$\gket\ transition is weakly allowed, with an Einstein A-coefficient of 4.9$\cdot$10$^5$ s$^{-1}$. In a high electric field, direct excitation from \gket\ to $M$-levels of \bket\ and \aket\ should thus become possible. 

We perform direct excitation from \gket\ to the opposite-parity doublet in electric fields up to 150 kV/cm. Using several mJ of pulsed, narrowband radiation of the PDA in a 2 mm diameter beam, propagating perpendicular to the atomic beam and polarized parallel to the applied electric field, we prepare electronically excited atoms in specific $M$-levels. Upon leaving the high electric field region, the atoms in $M$-levels that are down-shifted from $E_a$ go adiabatically over to the field-free \aket\ state and live long enough to reach the ionization detection region. Atoms can equally well be prepared in $M$-levels that are up-shifted from $E_b$, but this process is more difficult to observe as most of these atoms radiatively decay before reaching the detection region. We observe AC Stark-shifting and -broadening of lines when exciting from \gket\ to Stark-components that are up-shifted from \bket. This is attributed to the near-resonance of the frequency of the PDA with a transition from these levels to \dket\ (see Fig.~\ref{Fig_ExperimentalDetails}(b)).

\textcolor{black}{The excitation spectrum from \gket\ to \aket\ at large electric fields is shown in Fig.~\ref{Fig_StarkJ10}. The approximate 0.007-0.010 cm$^{-1}$ width of the lines in Fig.~\ref{Fig_StarkJ10} is mainly determined by the bandwidth of the excitation laser. As the electric field is increased, the M-components of \aket\ are more and more down-shifted with respect to their zero-field energy $E_a$. The frequency shifts are almost exclusively determined by the Stark-interaction of \aket\ and \bket and clearly show a linear Stark shift in the large electric field limit.
Using Eq. (1), we can unambiguously determine the $M$ quantum number for the observed spectral lines. In our electric field configuration, the lower the M-component, the larger the Stark shift. The induced electric dipole moment $\mu (M)$ for the $M$-levels is approaching the limiting value for $\mathcal{E} \rightarrow \infty$, i.e. $\mu (M) = \sqrt{(100-M^2)/3990} \cdot \mu_\text{TDM}$, and is above 1.0 D for $M \leq$4. }

The intensity of the lines in the spectrum shown in Fig.~\ref{Fig_StarkJ10} is almost exclusively determined by the Stark-interaction between \bket\ and \cket. This interaction mixes a fraction that is proportional to $M \cdot \mu^{\prime}_\text{TDM}\mathcal{E}/\Delta E^{\prime}$ of the wavefunction of \cket\ into \bket. As the laser polarization is parallel to the applied electric field, the $M$-dependence of the intensity of the electric dipole allowed \cket $\leftarrow$\gket\ transition is proportional to $(81 - M^2)$. Since \aket\ and \bket\ are nearly fully mixed, the intensity of the lines in the spectrum is expected to be proportional to $M^2\cdot(81-M^2)\cdot(\mu^{\prime}_\text{TDM}\mathcal{E}/\Delta E^{\prime})^2$, which explains the observed intensity pattern very well. By fitting the frequencies of spectral lines measured in electric fields from 90 - 150 kV/cm to a three-level Stark-interaction model, using the known values of $\Delta E$, $E_b$, and setting $\Delta E^{\prime}$ = 212 cm$^{-1}$, we find $\mu_\text{TDM}$ = 7.65 $\pm$ 0.05 D and $\mu^{\prime}_\text{TDM}$ = 7 $\pm$ 1 D (see Supplementary Information). We observe the effect of a differential static polarizability shift on the optical transitions in high electric fields. Its effect on the shift of the magnetic field insensitive microwave transition in weak electric fields is below 1 \% of the Stark effect.

In conclusion, we have characterized the opposite-parity doublet in the Dy atom around 17513 cm$^{-1}$ and experimentally confirm that an electric dipole moment comparable to that of polar molecules can be induced in these states. At an electric field of E = 1 kV/cm, the induced electric dipole moment of the M=0 level will be about 0.044 D, whereas the lifetime of this level in \aket\ is then reduced to about 100 ms.
\textcolor{black}{When the Dy atoms are polarized with an electric dipole moment in the range from 0.01 D up to 0.25 D, the product of the square of the induced electric dipole moment and the lifetime of the M-level in \aket\ is (in a very good approximation) constant and given by $2.0\cdot 10^{-6} (100-M^2)D^2\cdot s$.}
This opposite-parity doublet might be suited for investigating a variety of theoretical proposals that have so far been considered for paramagnetic molecules\cite{Micheli2006,Karra2016}. Interestingly, the Stark-coupling with a third level makes preparation of the Dy atoms in this doublet possible via a direct transition from the ground state. 

\begin{acknowledgments}
We acknowledge the technical support of Sebastian Kray, Marco De Pas and Russell Thomas as well as the contribution of Luca Diaconescu in the early stage of the experiments. G.V. acknowledges support from the European Union (ERC, LIRICO 101115996).
\end{acknowledgments}

\bibliography{biblio}

\end{document}